\documentclass[aps,prb,twocolumn,groupedaddress,10pt]{revtex4-1}
\usepackage{graphicx}
\begin{document}

\author{C.~L.~Richardson$^{1}$, S.~D.~Edkins$^{1}$, G.~R.~Berdiyorov$^{2}$, C.~J.~Chua$^{1}$, J.~P.~Griffiths$^{1}$, G.~A.~C.~Jones$^{1}$, M.~R.~Buitelaar$^{1,3,4}$, V. Narayan$^{1}$, F. Sfigakis$^{1}$, C.~G.~Smith$^1$, L.~Covaci$^{5}$, M.~R.~Connolly$^{1,6}$}

\affiliation{$^1$Cavendish Laboratory, Department of Physics, University of Cambridge, Cambridge CB3 0HE, UK}

\affiliation{$^2$Qatar Environment and Energy Research Institute, Qatar Foundation, P.O.\ Box 5825, Doha, Qatar}

\affiliation{$^3$Department of Physics and Astronomy, University College London, Gower Street, London WC1E 6BT, UK}

\affiliation{$^4$London Centre for Nanotechnology, University College London, Gordon Street, London WC1H 0AH, UK}

\affiliation{$^5$Departement Fysica, Universiteit Antwerpen, Groenenborgerlaan 171, 2020 Antwerpen, Belgium}

\affiliation{$^6$National Physical Laboratory, Hampton Road, Teddington TW11 0LW, UK}

\date{\today}

\title{Vortex detection and quantum transport in mesoscopic \\graphene Josephson junction arrays}

\begin{abstract}
We investigate mesoscopic Josephson junction arrays created by patterning superconducting disks on monolayer graphene, concentrating on the high-$T/T_c$ regime of these devices and the phenomena which contribute to the superconducting glass state in diffusive arrays. We observe features in the magnetoconductance at rational fractions of flux quanta per array unit cell, which we attribute to the formation of flux-quantized vortices. The applied fields at which the features occur are well described by Ginzburg-Landau simulations that take into account the number of unit cells in the array. We find that the mean conductance and universal conductance fluctuations are both enhanced below the critical temperature and field of the superconductor, with greater enhancement away from the graphene Dirac point.
\end{abstract}

\maketitle

\section{Introduction}
Superconducting vortices in a two-dimensional gas of chiral massless Dirac fermions are predicted to harbour excitations that resemble hypothetical elementary particles known as Majorana zero modes \cite{Jackiw1981, Fu2008}. Developing ways to create, control, and interfere vortices in candidate Dirac conductors---such as graphene and the surface states of three-dimensional topological insulators---is therefore important for detecting non-Abelian statistics and for implementing topologically protected quantum information processing \cite{Fu2009}.

In the absence of intrinsic pairing of electrons through the Cooper channel, attention has focused on using the proximity effect to induce a Dirac condensate in these materials \cite{Ghaemi2011}. In the proximity effect, a normal metal/superconductor interface generates phase-correlated quasiparticle pairs via Andreev reflection, which opens an effective gap in the normal metal. These pairs communicate phase information between superconductors in Josephson junctions, leading to supercurrent flow. Using graphene as the normal conductor also allows the normal-state resistance of the junction to be tuned via the carrier density \cite{Heersche2007, Kessler2010}. This leads to a superconducting-to-insulating transition in disordered arrays of Josephson junctions on graphene \cite{Allain2012}, and a superconducting-to-metallic transition with the addition of flux pinning in ordered arrays \cite{Han2014}. 

The presence of flux vortices in ordered two-dimensional arrays causes a phase frustration between superconducting islands \cite{Teitel1983}, which interacts with Andreev corrections to the conductance \cite{Klapwijk2004} and the normal-state quantum interference present in diffusive, substrate-supported graphene. The Josephson coupling between some pairs of phase-frustrated islands is enhanced due to quantum interference, leading to a superconducting glass state above the expected critical field at low temperatures \cite{Feigel'man2008, Han2014}.

In this paper we study how the conductance of monolayer graphene is modified in the presence of ordered arrays of superconducting disks, concentrating on the phenomena which contribute to the superconducting glass state in the high-$T/T_c$ regime of these devices. We detect the presence of proximity vortices in the graphene through changes in the magnetoconductance whenever the magnetic field generates an integer number of flux quanta through these Josephson junction arrays. We also find that, due to the proximity effect, both the mean conductance and universal conductance fluctuations are enhanced below the critical temperature and field of the superconductor, with greater enhancement away from the graphene Dirac point.

Device fabrication and experimental methods are detailed in Sec.\ II. Section III characterizes carrier density- and temperature-dependent transport through two devices with different disk diameters and separations, revealing an Andreev correction to the mean conductance. In Sec. IV, we discuss features in the magnetoconductance which are linked to the phase coherence length and array geometry of the devices. Section V further explores the subset of these features which are due to vortex pinning in the arrays (thus introducing phase frustration between disks) using Ginzburg-Landau simulations of the critical current. Finally, Sec. VI is a detailed study of the universal conductance fluctuations in one of the arrays as the leads and disks enter the superconducting state.

\begin{figure}[!t]
\includegraphics{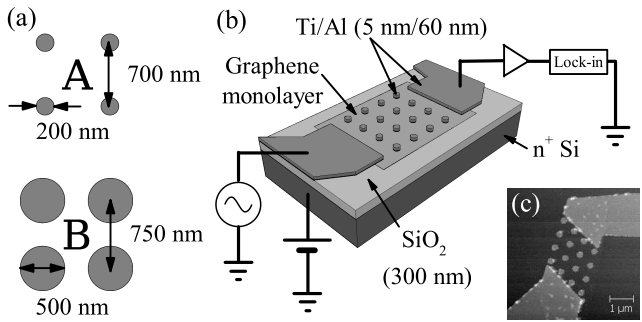}
\caption{(a) Schematics of array unit cells for device A (upper) and device B (lower). (b) Schematic of device A and two-terminal measurement circuit. (c) Atomic force microscope image of a similar device (disk diameter = 200 nm).}  
\label{Fig:Fig0}
\end{figure}

\section{Methods}
Our graphene flakes are mechanically exfoliated from natural graphite onto degenerately doped Si substrates with a 300 nm oxide layer. We identify monolayer flakes by optical contrast \cite{Bruna2009} and quantum Hall measurements \cite{Novoselov2005}, then pattern superconducting electrodes and square arrays of disks using electron beam lithography followed by thermal evaporation of a Ti/Al bilayer (5~nm/60~nm) [Fig.\ \ref{Fig:Fig0}]. We present measurements of two array geometries: device A has disks with diameter $2a$ = 200 nm and center-center separation $b$ = 700 nm, device B has disks with $2a$ = 500 nm and $b$ = 750 nm. The arrays have 3$\times$3 and 4$\times$3 unit cells, respectively.

Two-, four- and quasi-four-terminal differential conductance was measured using standard low frequency AC lock-in techniques, in a pumped $^3$He cryostat at 340~mK and a dilution refrigerator between 1.05~K and 100~mK. A voltage $V_{bg}$ applied to the doped Si substrate controlled the carrier density. For devices A and B, the graphene exhibited a Dirac point at $V_{bg}$ $\approx$ 2V and 6V and a carrier mobility of $\sim$3500~cm$^{2}$(Vs)$^{-1}$ and $\sim$5400~cm$^{2}$(Vs)$^{-1}$ for $n$~= 5~$\times$~10$^{11}$~cm$^{-2}$ and $T$~= 350~mK. The electron mean-free path $l_{e}=2D/v_F$ is approximately 66~nm for device A and 92~nm for device B, where $D$ is the diffusion constant ($\approx$~0.03~m$^2$s$^{-1}$ and 0.05~m$^2$s$^{-1}$).

\begin{figure}[!t]
\includegraphics{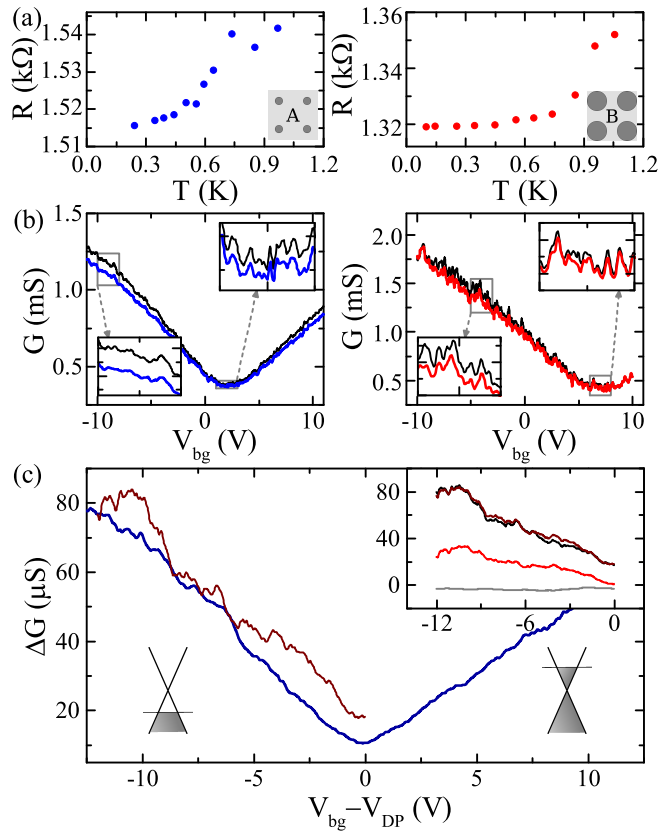}
\caption{(Color online) (a) Temperature dependence of the zero-field differential resistance in devices A (left) and B (right). (b) Back-gate voltage ($V_{bg}$) dependence of the differential conductance at zero field (thin black) and above the critical field of each device (thick blue/red, $B$~=~6~mT for device A and 12~mT for device B), $T$~=~350~mK. Grey boxes indicate data in insets. (c) Difference between the zero-field and critical-field conductance, $\Delta G$, as a function of back-gate voltage relative to the Dirac point, $V_{DP}$ (thick dark blue: device A, thin dark red: device B). Inset: Temperature dependence of $\Delta G$, device B. $T$~=~100~mK (black), 350~mK (dark red), 750~mK (red) and 1.05~K (grey).}
\label{Fig:Fig1}
\end{figure}

\section{Device characterization}
Figure \ref{Fig:Fig1}(a) shows the zero-field resistance of both devices as a function of temperature, corrected for lead resistance. We observe a decrease in both cases below the critical temperature of the Ti/Al bilayer, which is around 1~K. Figure \ref{Fig:Fig1}(b) plots the conductance of both devices above (thick red/blue lines) and below (thin black lines) the critical magnetic field as a function of back-gate voltage. As well as an increase of the overall conductance at zero field, we observe reproducible, aperiodic universal conductance fluctuations in both devices, the amplitude of which is also enhanced in the presence of the superconductor due to Andreev reflection at the interfaces \cite{Beenakker1997}, seen more clearly in the insets of Fig.\ \ref{Fig:Fig1}(b). See Sec.\ VI for a more detailed analysis and discussion of universal conductance fluctuation enhancement in device B.

We find that the mean conductance of both devices is larger at zero field [Fig.\ \ref{Fig:Fig1}(b)], with a greater difference between the superconducting and normal states as the carrier density increases [Fig.\ \ref{Fig:Fig1}(c)]. As the temperature is increased [inset of Fig.\ \ref{Fig:Fig1}(c)] the linear dependence on back-gate voltage persists, and by $T$ = 750 mK (red line) the magnitude of $\Delta G$ has approximately halved. Above the critical temperature of the leads, $\Delta G$ is small and negative due to suppression of weak localization by the small applied field \cite{McCann2006b}.

The data in Fig.\ \ref{Fig:Fig1} indicate that the devices behave as two-dimensional Josephson junction arrays, measured in the high-$T/T_c$ regime. Josephson junction arrays, unlike individual junctions, have an initial drop in resistance below the critical temperature of the superconductor but do not enter a zero-resistance state until a second (lower) critical temperature is reached \cite{Resnick1981}. Both critical temperatures are inversely proportional to the edge-edge separation of the disks in the array \cite{Eley2012}. In Fig.\ \ref{Fig:Fig1}(a), we see the initial resistance drop in both arrays, but not the transition to the zero-resistance state. The percentage decrease of the resistance is also not proportional to the percentage of the graphene covered by the disks (in contrast to Ref.\ \onlinecite{Han2014}), indicating a non-negligible interface resistance \cite{Blonder1982}. As expected, device B has a higher critical temperature, due to smaller edge-edge separation (250~nm compared to 500~nm in device A).

A non-uniform interface resistance could additionally lead to different parts of the disks opening an effective superconducting gap in the graphene at different threshold carrier densities \cite{Hammer2007}. Alternatively, disorder and the formation of charge puddles \cite{Martin2008} changes the phase coherence length in the graphene, and hence the extent of the proximity effect, as the carrier density is varied \cite{Komatsu2012}.

The absence of a zero-resistance state in these devices is most likely due to a combination of interface resistance and large edge-edge separation of the disks. Either could reduce the critical temperature of the zero-resistance state to below 250 or 100~mK (the lowest measured temperatures for devices A and B). However, without observing a supercurrent we cannot rule out destruction of the fully superconducting state by electromagnetic noise in the measurement circuit. In these measurements, we reduce low frequency noise using lock-in techniques and low-pass RC filters at room temperature, but the presence of unfiltered high-frequency photons may reduce the effective gap in the graphene \cite{Tinkham2004} and drive the devices into the high-$T/T_c$ regime.

\begin{figure}[!t]
\includegraphics{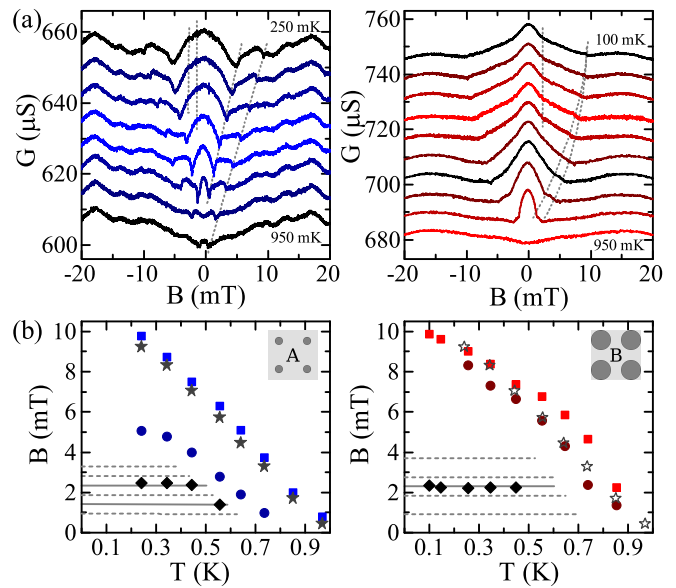}
\caption{(Color online) (a) Magnetoconductance at temperatures between 250~mK and 950~mK (100~mK and 150~mK traces also shown for device B). Traces offset for clarity, grey dashed lines highlight features plotted below. (b) Plots of the critical field of the Ti/Al bilayer (stars), outermost (squares) and innermost (circles) $T$-dependent features, and array-geometry--dependent features (diamonds). Grey horizontal lines are rational fractions of flux quanta per array unit cell; $\frac{2}{9}$ to $\frac{7}{9}$ in left panel, $\frac{1}{4}$ to $\frac{4}{4}$ and $\frac{5}{8}$ in right panel.}
\label{Fig:Fig2}
\end{figure}

\section{Magnetoconductance}
Having established that the modulation of the resistance is consistent with proximity effects around the islands, we now describe the magnetoconductance in greater detail. Figure \ref{Fig:Fig2}(a) shows the differential conductance captured as a function of magnetic field at different temperatures (bottom traces, $T$ = 950~mK; top traces, $T$ = 250~mK (device A) and 100~mK (device B)). Consistent with the results above, the application of a small magnetic field below $T_{c}$ decreases the conductance of both devices up to a critical field and for a wide range of carrier densities in both the electron and hole conduction regimes. Close inspection of these data reveals both temperature- and array-geometry--dependent features, indicated by grey dashed lines in Fig.\ \ref{Fig:Fig2}(a) and plotted in Fig.\ \ref{Fig:Fig2}(b). Also plotted, for reference, is the critical field of branched contacts on device A (grey stars). Note that sub-micron Al islands have been shown to have the same properties as continuous films \cite{Geim1997}, suggesting the disks have the same critical field and temperature as the leads.

The first set of $T$-dependent features closely follows the critical field of the Ti/Al bilayer in both devices. The second set appears at lower fields and has a similar temperature dependence, but differs in magnitude between devices, occurring at smaller fields and temperatures in device A. This correlation indicates that the two sets of features are related to the extent of the proximity effect in the graphene. 

Below the critical field of the disks, the onset of Andreev reflection at the interfaces changes the conditions for phase coherence across the device, increasing the global conductance \cite{Beenakker1997}. Once phase coherence is established between pairs of adjacent disks, the conductance of the graphene should be further enhanced \cite{Resnick1981, Eley2012}. We observe this as the second set of features. Estimates of the phase coherence length from the weak localization peak seen at 950 mK [bottom traces in Fig. 3(a)] are approximately 200 nm for both devices. Therefore, as the temperature decreases, phase coherence develops between the disks in device B before device A, and the features are seen at higher temperatures in device B.

Below the second $T$-dependent feature, we observe features at fields related to the area of the array unit cell. These fields are indicated by grey horizontal lines in Fig.\ \ref{Fig:Fig2}(b). Such features at fixed, geometry-dependent fields are expected in two-dimensional Josephson junction arrays \cite{Teitel1983, Fiory1978} due to low-energy configurations of vortices within the array, such as integer numbers of flux quanta per unit cell, \textit{f}, or a superlattice of empty and occupied cells. We observe geometry-dependent features at $B$ = 1.38 mT, and 2.37--2.46 mT for device A (close to $f = \frac{1}{3}$ = 1.41 mT and $\frac{5}{9}$ = 2.35 mT), and at $B$ = 2.20--2.33 mT for device B ($f = \frac{5}{8}$ = 2.30 mT). In a large, square array we expect the lowest energy configurations at fields of $\frac{n}{4}$ flux quanta per unit cell \cite{VanDerZant1994}.

\begin{figure}[!b]
\includegraphics{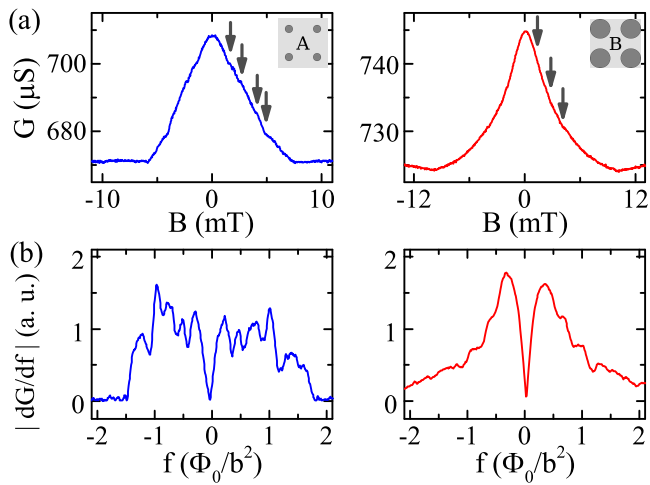}
\caption{(Color online) (a) Differential conductance of device A (left) and B (right) as a function of magnetic field, averaged over back-gate voltage. $T$~=~350~mK. (b) Absolute derivative of data in (a) plotted as a function of flux quanta per array unit cell.}
\label{Fig:Fig5}
\end{figure}

\section{Vortex pinning in the arrays}
\begin{figure}
\includegraphics{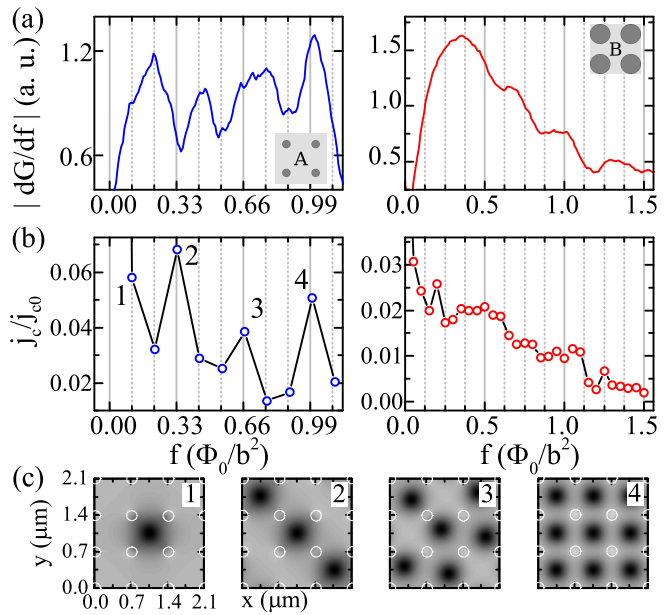}
\caption{(Color online) (a) Absolute derivative of the back-gate--averaged conductance of devices A (left) and B (right) as a function of flux quanta per array unit cell, \textit{f}. Grey solid and dashed lines indicate rational fractional values of \textit{f}~=~$\frac{n}{9}$ for device A and $\frac{n}{8}$ for device B. (b) Calculated critical current through the arrays as a function of \textit{f}, normalized to the zero-field value, using the Ginzburg-Landau formalism. (c) Local magnetic field distribution in device A (black = high field, grey = low field) at applied fields indicated by numbered points in left panel of (b). White circles show disk locations. Ginzburg-Landau results are obtained at $T$~=~0.9$T_{c}$ for anisotropy coefficient $g(r)$~=~0.25, coherence length $\xi$(0)~=~20~nm, and penetration depth $\lambda$(0)~=~200~nm.}
\label{Fig:Fig3}
\end{figure}

The fields at which ground-state vortex pinning configurations occur should depend only on the array geometry \cite{Teitel1983}, not temperature [as we see in Fig.\ \ref{Fig:Fig2}] or normal state conductance \cite{VanDerZant1992a, VanDerZant1994}. We confirm that these geometry-dependent features are due to vortex pinning in the arrays by averaging the magnetoconductance over a 20~V range of back-gate voltage [shown in Fig.\ \ref{Fig:Fig5}(a)] and differentiating [the absolute derivative is plotted in Fig.\ \ref{Fig:Fig5}(b)]. Averaging over back-gate voltage eliminates both universal conductance fluctuations and any normal state conductance dependence. The absolute value of the derivative is plotted again in Fig.\ \ref{Fig:Fig3}(a) with gridlines at fractions of flux quanta per unit cell of $\frac{n}{9}$ for device A, and $\frac{n}{8}$ for device B. We find features approximately at fractions of $\frac{n}{9}$ for device A, and around $\frac{3}{8}$, $\frac{5}{8}$, $\frac{8}{8}$ and $\sim \frac{10}{8}$ for device B, strongly suggesting that the features are linked to the unit cell area of the arrays and are independent of the normal-state resistance.

To understand why the features occur at these fields rather than at \textit{f}~=~$\frac{n}{4}$, we perform simulations based on the Ginzburg-Landau equations. The model systems consist of square arrays of superconducting disks embedded in a normal metal matrix, with the same geometry as our devices (including the voltage probes of device B). The proximity effect is described by a spatially-dependent, anisotropic expansion coefficient of the Gibbs free energy functional, $\alpha= \alpha_0 g(r)$, which equals 1 inside the superconducting disks and less than 1 in the metallic regions \cite{Berdiyorov2012}. The square simulation region has periodic boundary conditions in all directions in the two-dimensional plane, and is exposed to a homogeneous magnetic field $B$. Critical current, plotted in Fig.\ \ref{Fig:Fig3}(b) as a function of \textit{f}, is calculated by slowly driving current through the ground state vortex configuration of the system \cite{Berdiyorov2006b}. The configurations are obtained in field-cooled simulations \cite{Berdiyorov2006a}, since the initial vortex structure directly influences the response of the system to an applied current \cite{Latimer2012}. Figure~\ref{Fig:Fig3}(c) shows simulated local magnetic field density in device A for the four applied fields indicated in Fig.\ \ref{Fig:Fig3}(b) (white circles indicate the location of the superconducting disks). The formation of vortices in the array, and of superlattices at values of \textit{f}~$\neq \frac{n}{4}$, is demonstrated.

We measure our arrays in the high-$T/T_c$ regime, below their initial transition temperature but above the (Berezinskii-Kosterlitz-Thouless) transition to the zero-resistance state. It appears that in this regime, the changes in flux-flow resistance due to vortex pinning and motion \cite{Rzchowski1990} are visible in addition to the larger resistance caused by dissociation of vortex-antivortex pairs \cite{Resnick1981}. Therefore, taking the derivative of the differential conductance reveals the changes in the flux-flow resistance. We note that Han \emph{et al.} \cite{Han2014} also measure similar features in the differential conductance above the critical temperature of the zero-resistance state, supporting our interpretation of the differential conductance being measured in the high-$T/T_c$ regime of the arrays.

Since the critical current of an array is the current at which vortices are depinned, we expect it to be proportional to the differential conductance (critical current is directly proportional to the differential resistance in the fully superconducting state \cite{VanDerZantThesis}). In fact, we do see similar trends and features in the experimental data and simulated critical current [Fig.\ \ref{Fig:Fig3}]. For device A, the sharp features appearing at \textit{f}~$\approx \frac{n}{3}$ are reflected in the critical current. For device B, we note the similarity between the smooth overall trends and broad peaks at \textit{f}~$\approx \frac{5}{8}$ and $\frac{10}{8}$ in simulation and experiment. 

These correlations indicate that the small number of unit cells does indeed give rise to the observed position of the features in field. However, there are differences between the simulations and experiments, for example the number of features seen in device A and whether they are maxima or minima. The exact relationship between critical current in the zero-resistance state and flux-flow resistance near $T_c$ in Josephson junction arrays is not clear, and could account for the differences. Additionally, the simulations do not include the disorder potential in the graphene, which would alter the pinning potential seen by vortices as well as introduce quantum interference corrections to the conductance. 

\section{Universal conductance fluctuations}
Along with the vortex-induced phase frustration effects detailed above, the presence of reproducible, aperiodic conductance fluctuations in our mesoscale flakes is a precursor of the superconducting glass state predicted \cite{Feigel'man2008} and recently observed \cite{Han2014} in ordered graphene Josephson junction arrays. Here, `glass' refers to a full proximity effect above the expected critical field, in the case where the superconducting phase of each disk is randomly oriented---ordinarily, the phases of the islands in an array are identical in the zero-resistance state. In an array with a diffusive normal metal, quantum interference can enhance Josephson coupling between some pairs of disks enough for supercurrent to flow.

\begin{figure}
\includegraphics{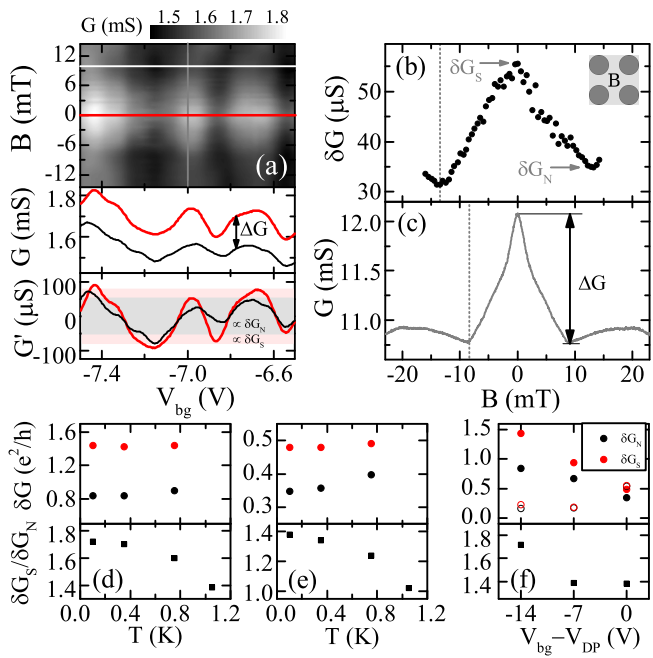}
\caption{(Color online) $T$ = 100 mK, $V_{bg}-V_{DP}$ = -14V unless stated otherwise ($V_{DP}$ = back-gate voltage corresponding to Dirac point). (a) Differential conductance of device B as a function of back-gate voltage and magnetic field (thick red lines: $B=0$, thin black lines: $B=10$ mT, $G'$ is the detrended conductance). (b) Standard deviation of universal conductance fluctuations as a function of magnetic field; $\delta G_{N}$ and $\delta G_{S}$ are the values at $B=10$ and 0~mT. (c) Differential conductance as a function of magnetic field (d),~(e) Upper panel: Temperature dependence of $\delta G_{N}$ (black) and $\delta G_{S}$ (red) in terms of conductance quanta at $V_{bg}-V_{DP}$ = -14V and 0V, respectively. Lower panel: Ratio of $\delta G_{S}$ and $\delta G_{N}$. (f) As for (d) and (e), but showing carrier density dependence at 100~mK (solid markers) and 1.05~K (hollow markers).}
\label{Fig:Fig4}
\end{figure}

In Fig.\ \ref{Fig:Fig4} we explore the effect of the superconducting disks on universal conductance fluctuations (UCFs). Figure \ref{Fig:Fig4}(a) shows UCFs in device B as a function of back-gate voltage and applied field, at $T$ = 100 mK and $V_{bg}-V_{DP}$ = -14V (where $V_{DP}$ is the back-gate voltage corresponding to the Dirac point). The overall conductance of the array is enhanced below the critical field of the leads (middle panel), however, the UCFs themselves are also enhanced, as shown in the lower panel (the mean conductance has been subtracted to give the detrended conductance, $G'$). Figure \ref{Fig:Fig4}(b) shows the standard deviation of the UCFs, $\delta G = \sqrt{\langle G^{2} \rangle - \langle G \rangle^{2}}$, as a function of magnetic field. Below the critical field of the leads, $\delta G$ increases linearly in magnitude towards zero-field. We observe this linear trend at different carrier densities and temperatures. Comparison of $\delta G$ with the differential conductance of the device [Fig.\ \ref{Fig:Fig4}(c)] reveals that $\delta G$ is enhanced over a larger field range than the mean conductance. 

Figure \ref{Fig:Fig4}(d)-(f) quantifies the enhancement of $\delta G$ between $B=0$ ($\delta G_{S}$) and $B=10$ mT ($\delta G_{N}$), and compares data taken at different back-gate voltages and temperatures. Upper panels show the values of $\delta G_{S}$ and $\delta G_{N}$, and lower panels their ratio. The form of the temperature dependence of this ratio is the same at both high and low carrier density in Fig.\ \ref{Fig:Fig4}(d) and (e). Figure \ref{Fig:Fig4}(f) shows the back-gate voltage dependence at 100~mK, and above the critical temperature of the leads at 1.05~K. The enhancement of $\sim$1.4--1.7 is close to that observed in InAs nanowires \cite{Doh2008} and graphene \cite{Trbovic2010} contacted by Ti/Al bilayers. At 100~mK, $\delta G$ is more enhanced away from the Dirac point, whereas at 1.05~K we observe the opposite trend in $\delta G_{N}$, with no enhancement at zero-field.

The enhancement of UCFs in mesoscopic devices with superconducting contacts is associated with Andreev reflection at the normal/superconductor interface \cite{Beenakker1997}. The temperature dependence of the UCF enhancement, and its onset at larger applied fields compared to the mean conductance, can be linked to the superconducting gap in the Ti/Al bilayer. A larger gap leads to more carriers being Andreev reflected at the interfaces, and therefore greater enhancement of the UCFs. Our observation of a smaller enhancement close to the Dirac point is more intriguing, since suppression of superconductivity near the graphene Dirac point has been seen before in large-area devices \cite{Komatsu2012, Han2014}. So far, the suppression has only been seen in the global conductance, but it also appears here in conductance fluctuation data. Loss of phase coherence caused by the formation of electron-hole puddles at low carrier densities \cite{Martin2008} appears to be responsible for the suppression in both global conductance and fluctuations \cite{Komatsu2012}. A simple normal-state resistance dependence is unlikely, since the ratio of $\delta G_{S}$ and $\delta G_{N}$ is the same at $V_{bg}-V_{DP}$ = -7V and the Dirac point. Another possibility is the loss of Andreev pairs due to specular Andreev reflection at the graphene/disk interfaces \cite{Beenakker2006}, though a measurable contribution is unlikely given the diffusive transport in our substrate-supported graphene.

\section{Conclusion}
In summary, we have explored charge transport in the high-$T/T_c$ regime of ordered arrays of mesoscopic superconductors on graphene. Even outside the zero-resistance state, magnetoconductance features at rational fractions of flux quanta per array unit cell indicate the presence of proximity vortices. Ginzburg-Landau simulations capture the effects of a small number of unit cells on the magnetoconductance. We also observe suppression of the proximity effect at the Dirac point through the mean conductance and universal conductance fluctuations, similar to that observed in other large-area, diffusive graphene proximity devices \cite{Han2014, Komatsu2012}.

These measurements of graphene-based Josephson junction arrays demonstrate the feasibility of detecting and manipulating superconducting vortices in Dirac conductors through charge transport. Extension of this approach to topological insulators could be a route to detection of non-Abelian statistics and implementation of topological quantum computing.

\begin{acknowledgments}
This work was financially supported by the Engineering and Physical Sciences Research Council, and an NPL/EPSRC Joint Postdoctoral Partnership. Data related to this publication is available at the DSpace@Cambridge data repository (https://www.repository.cam.ac.uk/handle/1810/248242).
\end{acknowledgments}


\begin{thebibliography}{35}
\providecommand{\natexlab}[1]{#1}
\providecommand{\url}[1]{\texttt{#1}}
\expandafter\ifx\csname urlstyle\endcsname\relax
  \providecommand{\doi}[1]{doi: #1}\else
  \providecommand{\doi}{doi: \begingroup \urlstyle{rm}\Url}\fi

\bibitem[Jackiw and Rossi(1981)]{Jackiw1981}
R.~Jackiw and P.~Rossi.
\newblock \emph{Nucl. Phys. B}, {\bf 190}, 681 -- 691 (1981).

\bibitem[Fu and Kane(2008)]{Fu2008}
Liang Fu and C.~L. Kane.
\newblock \emph{Phys. Rev. Lett.}, {\bf 100}, 096407 (2008).

\bibitem[Fu and Kane(2009)]{Fu2009}
Liang Fu and C.~L. Kane.
\newblock \emph{Phys. Rev. B}, {\bf 79}, 161408 (2009).

\bibitem[Ghaemi and Wilczek(2012)]{Ghaemi2011}
P.~Ghaemi and F.~Wilczek.
\newblock \emph{Phys. Scripta}, {\bf 2012} (T146), 014019 (2012).

\bibitem[{Heersche} et~al.(2007){Heersche}, {Jarillo-Herrero}, {Oostinga},
  {Vandersypen}, and {Morpurgo}]{Heersche2007}
H.~B. {Heersche}, P.~{Jarillo-Herrero}, J.~B. {Oostinga}, L.~M.~K.
  {Vandersypen}, and A.~F. {Morpurgo}.
\newblock \emph{Nature}, {\bf 446}, 56--59 (2007).

\bibitem[Kessler et~al.(2010)Kessler, Girit, Zettl, and Bouchiat]{Kessler2010}
B.~M. Kessler, \ifmmode \mbox{\c{C}}\else \c{C}\fi{}.~\"O. Girit, A.~Zettl, and
  V.~Bouchiat.
\newblock \emph{Phys. Rev. Lett.}, {\bf 104}, 047001 (2010).

\bibitem[Allain et~al.(2012)Allain, Han, and Bouchiat]{Allain2012}
A.~Allain, Z.~Han, and V.~Bouchiat.
\newblock \emph{Nat. Mater.}, {\bf 11}, 190--194 (2012).

\bibitem[Han et~al.(2014)Han, Allain, Arjmandi-Tash, Tikhonov, Feigel'man,
  Sac\'{e}p\'{e}, and Bouchiat]{Han2014}
Z.~Han, A.~Allain, H.~Arjmandi-Tash, K.~Tikhonov, M.~Feigel'man,
  B.~Sac\'{e}p\'{e}, and V.~Bouchiat.
\newblock \emph{Nat. Phys.}, {\bf 10}, 380--386 (2014).

\bibitem[Teitel and Jayaprakash(1983)]{Teitel1983}
S.~Teitel and C.~Jayaprakash.
\newblock \emph{Phys. Rev. Lett.}, {\bf 51}, 1999--2002 (1983).

\bibitem[Klapwijk(2004)]{Klapwijk2004}
T.~M. Klapwijk.
\newblock \emph{J. Supercond. Nov. Magn.}, {\bf 17}, 593--611 (2004).

\bibitem[{Feigel'Man} et~al.(2008){Feigel'Man}, {Skvortsov}, and
  {Tikhonov}]{Feigel'man2008}
M.~V. {Feigel'Man}, M.~A. {Skvortsov}, and K.~S. {Tikhonov}.
\newblock \emph{JETP Lett.}, {\bf 88}, 747--751 (2008).

\bibitem[Bruna and Borini(2009)]{Bruna2009}
Matteo Bruna and Stefano Borini.
\newblock \emph{J. Phys. D: Appl. Phys.}, {\bf 42}, 175307 (2009).

\bibitem[{Novoselov} et~al.(2005){Novoselov}, {Geim}, {Morozov}, {Jiang},
  {Katsnelson}, {Grigorieva}, {Dubonos}, and {Firsov}]{Novoselov2005}
K.~S. {Novoselov}, A.~K. {Geim}, S.~V. {Morozov}, D.~{Jiang}, M.~I.
  {Katsnelson}, I.~V. {Grigorieva}, S.~V. {Dubonos}, and A.~A. {Firsov}.
\newblock \emph{Nature}, {\bf 438}, 197--200 (2005).

\bibitem[Beenakker(1997)]{Beenakker1997}
C.~W.~J. Beenakker.
\newblock \emph{Rev. Mod. Phys.}, {\bf 69}, 731--808 (1997).

\bibitem[McCann et~al.(2006)McCann, Kechedzhi, Fal'ko, Suzuura, Ando, and
  Altshuler]{McCann2006b}
E.~McCann, K.~Kechedzhi, Vladimir~I. Fal'ko, H.~Suzuura, T.~Ando, and B.~L.
  Altshuler.
\newblock \emph{Phys. Rev. Lett.}, {\bf 97}, 146805 (2006).

\bibitem[Resnick et~al.(1981)Resnick, Garland, Boyd, Shoemaker, and
  Newrock]{Resnick1981}
D.~J. Resnick, J.~C. Garland, J.~T. Boyd, S.~Shoemaker, and R.~S. Newrock.
\newblock \emph{Phys. Rev. Lett.}, {\bf 47}, 1542--1545 (1981).

\bibitem[Eley et~al.(2012)Eley, Gopalakrishnan, Goldbart, and Mason]{Eley2012}
S.\ Eley, S.\ Gopalakrishnan, P.\~M.\ Goldbart, and N.\ Mason.
\newblock \emph{Nat. Phys.}, {\bf 8}, 59--62 (2012).

\bibitem[Blonder et~al.(1982)Blonder, Tinkham, and Klapwijk]{Blonder1982}
G.~E. Blonder, M.~Tinkham, and T.~M. Klapwijk.
\newblock \emph{Phys. Rev. B}, {\bf 25}, 4515--4532 (1982).

\bibitem[Hammer et~al.(2007)Hammer, Cuevas, Bergeret, and Belzig]{Hammer2007}
J.~C. Hammer, J.~C. Cuevas, F.~S. Bergeret, and W.~Belzig.
\newblock \emph{Phys. Rev. B}, {\bf 76}, 064514 (2007).

\bibitem[{Martin} et~al.(2008){Martin}, {Akerman}, {Ulbricht}, {Lohmann},
  {Smet}, {von Klitzing}, and {Yacoby}]{Martin2008}
J.~{Martin}, N.~{Akerman}, G.~{Ulbricht}, T.~{Lohmann}, J.~H. {Smet}, K.~{von
  Klitzing}, and A.~{Yacoby}.
\newblock \emph{Nat. Phys.}, {\bf 4}, 144--148 (2008).

\bibitem[Komatsu et~al.(2012)Komatsu, Li, Autier-Laurent, Bouchiat, and
  Gu\'eron]{Komatsu2012}
Katsuyoshi Komatsu, Chuan Li, S.~Autier-Laurent, H.~Bouchiat, and S.~Gu\'eron.
\newblock \emph{Phys. Rev. B}, {\bf 86}, 115412 (2012).

\bibitem[Tinkham(2004)]{Tinkham2004}
M.~Tinkham.
\newblock \emph{Introduction to Superconductivity}.
\newblock Dover Publications, 2nd ed. edition (2004).

\bibitem[{Geim} et~al.(1997){Geim}, {Grigorieva}, {Dubonos}, {Lok}, {Maan},
  {Filippov}, and {Peeters}]{Geim1997}
A.~K. {Geim}, I.~V. {Grigorieva}, S.~V. {Dubonos}, J.~G.~S. {Lok}, J.~C.
  {Maan}, A.~E. {Filippov}, and F.~M. {Peeters}.
\newblock \emph{Nature}, {\bf 390}, 259--262 (1997).

\bibitem[Fiory et~al.(1978)Fiory, Hebard, and Somekh]{Fiory1978}
A.~T. Fiory, A.~F. Hebard, and S.~Somekh.
\newblock \emph{Appl. Phys. Lett.}, {\bf 32}, 73--75 (1978).

\bibitem[van~der Zant et~al.(1994)van~der Zant, Webster, Romijn, and
  Mooij]{VanDerZant1994}
H.~S.~J. van~der Zant, M.~N. Webster, J.~Romijn, and J.~E. Mooij.
\newblock \emph{Phys. Rev. B}, {\bf 50}, 340--350 (1994).

\bibitem[van~der Zant et~al.(1992)van~der Zant, Fritschy, Elion, Geerligs, and
  Mooij]{VanDerZant1992a}
H.~S.~J. van~der Zant, F.~C. Fritschy, W.~J. Elion, L.~J. Geerligs, and J.~E.
  Mooij.
\newblock \emph{Phys. Rev. Lett.}, {\bf 69}, 2971--2974 (1992).

\bibitem[Berdiyorov et~al.(2012)Berdiyorov, Romaguera, Milo\v{s}evi\'{c}, Doria,
  Covaci, and Peeters]{Berdiyorov2012}
G.R. Berdiyorov, A.R. Romaguera, M.V. Milo\v{s}evi\'{c}, M.M. Doria, L.~Covaci, and
  F.M. Peeters.
\newblock \emph{Eur. Phys. J. B}, {\bf 85}, 1--8 (2012).

\bibitem[Berdiyorov et~al.(2006{\natexlab{a}})Berdiyorov, Milo\v{s}evi\'{c},
  and Peeters]{Berdiyorov2006b}
G.~R. Berdiyorov, M.~V. Milo\v{s}evi\'{c}, and F.~M. Peeters.
\newblock \emph{Europhys. Lett.}, {\bf 74}, 493 (2006){\natexlab{a}}.

\bibitem[Berdiyorov et~al.(2006{\natexlab{b}})Berdiyorov, Milo\v{s}evi\'{c},
  and Peeters]{Berdiyorov2006a}
G.~R. Berdiyorov, M.~V. Milo\v{s}evi\'{c}, and F.~M. Peeters.
\newblock \emph{Phys. Rev. Lett.}, {\bf 96}, 207001 (2006){\natexlab{b}}.

\bibitem[Latimer et~al.(2012)Latimer, Berdiyorov, Xiao, Kwok, and
  Peeters]{Latimer2012}
M.~L. Latimer, G.~R. Berdiyorov, Z.~L. Xiao, W.~K. Kwok, and F.~M. Peeters.
\newblock \emph{Phys. Rev. B}, {\bf 85}, 012505 (2012).

\bibitem[Rzchowski et~al.(1990)Rzchowski, Benz, Tinkham, and
  Lobb]{Rzchowski1990}
M.~S. Rzchowski, S.~P. Benz, M.~Tinkham, and C.~J. Lobb.
\newblock \emph{Phys. Rev. B}, {\bf 42}, 2041--2050 (1990).

\bibitem[van~der Zant(1991)]{VanDerZantThesis}
H.~S.~J. van~der Zant.
\newblock PhD thesis, Technische Universiteit Delft (1991).

\bibitem[Doh et~al.(2008)Doh, Franceschi, Bakkers, and Kouwenhoven]{Doh2008}
Yong-Joo Doh, Silvano~De Franceschi, Erik P. A.~M. Bakkers, and Leo~P.
  Kouwenhoven.
\newblock \emph{Nano Lett.}, {\bf 8}, 4098--4102 (2008).

\bibitem[{Trbovic} et~al.(2010){Trbovic}, {Minder}, {Freitag}, and
  {Sch{\"o}nenberger}]{Trbovic2010}
J.~{Trbovic}, N.~{Minder}, F.~{Freitag}, and C.~{Sch{\"o}nenberger}.
\newblock \emph{Nanotechnology}, {\bf 21}, A264005 (2010).

\bibitem[Beenakker(2006)]{Beenakker2006}
C.~W.~J. Beenakker.
\newblock \emph{Phys. Rev. Lett.}, {\bf 97}, 067007 (2006).

\end{thebibliography}
\end{document}